\documentclass[final,3p,twocolumn]{elsarticle}

\usepackage{graphicx}
\usepackage{amsmath, amsfonts, amssymb, bm}

\usepackage{amstext}
\usepackage{color}

\begin{document}

\title{Investigation of classical radiation reaction with aligned crystals}

\author{A. Di Piazza}

\address{Max-Planck-Institut f\"{u}r Kernphysik, Saupfercheckweg 1, D-69117, Germany}

\author{T. N. Wistisen, U. I. Uggerh\o j}

\address{Department of Physics and Astronomy, Aarhus University, 8000 Aarhus,
Denmark\\ 
\vspace{1cm}
This article is registered under preprint number: arXiv:1503.05717}

\begin{abstract}
Classical radiation reaction is the effect of the electromagnetic field emitted by an accelerated
electric charge on the motion of the charge itself. The self-consistent
underlying classical equation of motion including radiation-reaction effects, the Landau-Lifshitz
equation, has never been tested experimentally, in spite of the first theoretical 
treatments of radiation reaction having been developed more than a century ago. Here we show that classical
radiation reaction effects, in particular those due to the near electromagnetic field, 
as predicted by the Landau-Lifshitz equation, can be measured in principle using presently available facilities, 
in the energy emission spectrum of $30\text{-}\text{GeV}$ electrons crossing a $0.55$-$\text{mm}$ thick diamond 
crystal in the axial channeling regime. Our theoretical results indicate the 
feasibility of the suggested setup, e.g., at the CERN Secondary Beam Areas (SBA) beamlines.

\vspace{1cm}
\end{abstract}

\maketitle

\section{Introduction}
The Lorentz equation is one of the cornerstones of classical electrodynamics and
it describes the motion of an electric charge, an electron for definiteness (charge $e<0$ and mass $m$),
in the presence of an external, given electromagnetic field \cite{Jackson_b_1975}. The Lorentz equation,
however, does not take into account that, as the electron is being accelerated
by the external field, it emits electromagnetic radiation, which in turn
alters the trajectory of the electron itself (radiation reaction (RR)).
The search for the equation of motion of an electron moving
in a given external electromagnetic field, including self-consistently
the effects of RR, has already been pursued since the beginning of the 20th century. 
By starting from the Lorentz equation of an electron in the
presence of an external electromagnetic field and of the electromagnetic field 
produced by the electron itself, the so-called Lorentz-Abraham-Dirac
(LAD) equation has been derived \cite{Abraham_b_1905,Lorentz_b_1909,Dirac_1938,
Jackson_b_1975,Landau_b_2_1975,Barut_b_1980,Hartemann_b_2001,Rohrlich_b_2007}.
After mass renormalization RR effects result in two force terms in the LAD
equation, one proportional to the Li\'{e}nard formula for the radiated power
and accounting for the energy-momentum loss of the electron due to radiation, the ``damping
term'', and the other one, the ``Schott'' term, related to the electron's near field \cite{Rohrlich_b_2007}
and accounting for the work done by the field emitted by the electron on the 
electron itself \cite{Hammond_2010}. Unlike the damping term, 
the Schott term, being proportional to the time derivative of the acceleration of the electron, 
1) renders the LAD equation a non-Newtonian, third-order time differential equation; and
2) allows for unphysical features of the LAD equation as the existence of 
``runaway solutions'', with the electron acceleration 
exponentially diverging in the remote future, even if, for example,
the external field identically vanishes 
\cite{Jackson_b_1975,Landau_b_2_1975,Barut_b_1980,Hartemann_b_2001,Rohrlich_b_2007,Hammond_2010,Di_Piazza_2012,Burton_2014}.

The origin of the inconsistencies of the LAD equation has been identified 
in \cite{Landau_b_2_1975}. The conclusion is that in the realm of classical electrodynamics, i.e., when quantum
effects can be neglected, a ``reduction of order'' can be consistently carried out in the LAD
equation, resulting in a second-order differential equation,
known as the Landau-Lifshitz (LL) equation. Moreover, quoting Spohn \cite{Spohn_2000},
the physical solutions of the LAD equation ``are on the critical manifold and are
governed there by an effective second-order equation'' which is the LL
equation. Finally, the LL equation has been also derived from quantum electrodynamics 
in \cite{Krivitskii_1991} (see also \cite{Ilderton_2013}).

The rapid progress of laser technology has renewed the interest in the problem of
RR as the strong electromagnetic fields produced
by lasers can violently accelerate the electron and 
consequently prime a substantial emission of electromagnetic
radiation. Correspondingly, a large number of setups and
schemes have been recently proposed to measure classical 
RR effects in electron-laser interaction \cite{Thomas_2012,Gonoskov_2013,Kumar_2013,Tamburini_2014,Zhidkov_2014,Heinzl_2015} 
(we refer to the review \cite{Di_Piazza_2012} for previous proposals). However,
experimental challenges either in the detection of relatively small RR effects or in 
the availability of sufficiently strong lasers has prevented so far any 
experimental test of the LL equation. Moreover, since RR effects are larger 
for ultrarelativistic electrons, reported laser-based experimental
tests of the LL equation turn out to be sensitive mainly to the damping term in the LL equation,
which has the most favorable dependence on the electron Lorentz factor.

In the present Letter we adopt a different perspective
and put forward a presently feasible experimental setup to
measure classical RR effects on the radiation field,
generated in the interaction of ultrarelativistic electrons with an aligned crystal.
The experiment can already be performed at, e.g., the CERN Secondary Beam Areas (SBA) beamlines.
In fact, in the proposed setup $30\text{-}\text{GeV}$ electrons impinge 
into a $0.55\text{-}\text{mm}$ thick diamond crystal and emit a significant
amount of radiation due to axial channeling \cite{Artru_1994,Akhiezer_b_1996,Baier_b_1998,Uggerhoj_2005}. 
Our numerical simulations indicate that in this regime RR effects substantially alter 
the electromagnetic emission spectrum. Moreover, unlike experimental proposals employing lasers, 
the distinct structure of the electric field of the crystal
at axial channeling renders the emission spectrum more sensitive
to a term in the LL equation originating from the 
controversial Schott term in the LAD equation. As we will see below, 
this term depends in general on the spacetime derivatives of the background field. 
This feature makes our setup prominent also with respect to synchrotron facilities
where the electron dynamics is dominated by the damping term. We also mention 
that at an electron energy $\varepsilon_0=30\;\text{GeV}$ and for a typical synchrotron radius
$R=1\;\text{km}$, the relative electron energy loss per turn is 
$\Delta\varepsilon/\varepsilon_0=8.9\times 10^{-5}\varepsilon_0[\text{GeV}]^3/R[\text{m}]=2.4\times 10^{-3}$ \cite{Wiedemann_b_2003}, which would induce too small effects on the emitted radiation to be measured. In addition, in order
for the synchrotron to operate during many turns, the electron energy loss 
has to be precisely compensated preventing again 
any possibility of ``accumulating'' and measuring RR effects on the emitted radiation.

\section{The physical model}
When a high-energy electron impinges onto a single crystal along a direction of
high symmetry, its motion can become transversely bound and its dynamics 
determined by a coherent scattering in the collective, screened
field of many atoms aligned along the direction of symmetry (axial channeling)
\cite{Artru_1994,Akhiezer_b_1996,Baier_b_1998,Uggerhoj_2005}. In this regime 
the electron experiences an effective potential in the transverse directions (continuum potential),
resulting from the average of the atomic potential along the direction of symmetry. 
For the sake of simplicity, in the present and in the next section we assume that the atomic
potential is due to a single string.
By indicating as $\bm{z}$ the direction corresponding to the symmetry axis 
of the crystal and by $\bm{\rho}=(x,y)$ the coordinates in the
transverse plane, with the atomic string crossing this plane 
at $\bm{\rho}=\bm{0}$, the continuum potential $\Phi(\rho)$ depends
only on the distance $\rho=|\bm{\rho}|$ and it can be approximated 
as \cite{Baier_b_1998}: 
\begin{equation}
\Phi(\rho)=
\Phi_{0}\left[\ln\left(1+\frac{1}{\varrho^2+\eta}\right)-\ln\left(1+\frac{1}{\varrho_c^2+\eta}\right)\right],\label{eq:potential}
\end{equation}
where $\bm{\varrho}=\bm{\rho}/a_s$ and $\varrho_c=\rho_c/a_s$. Here, the parameters $\Phi_0$, $\rho_c$, $\eta$, and $a_s$ depend on the crystal and $\rho\le\rho_c$. A convenient choice to investigate classical RR effects is diamond, with, e.g., the $\langle111\rangle$ as symmetry axis and for which $\Phi_0=29\;\text{V}$, $\rho_c=0.765\;\text{\AA}$, $\eta=0.025$, and $a_s=0.326\;\text{\AA}$. In fact, the relatively low value of $\Phi_0$ as compared to other crystals allows one to neglect quantum effects also at relatively high electron energies. The depth $\Phi_M=\Phi(0)$ of the potential in diamond is such that $U_M=U(0)=-103\;\text{eV}$, where $U(\rho)=e\Phi(\rho)$ is the electron potential energy (units with $\hbar=c=1$ and $\alpha=e^2\approx 1/137$ are employed throughout). 

In general, the channeling regime of interaction features
ultrastrong electromagnetic fields, which can lead to substantial 
energy loss of the radiating electron. In order for quantum effects to be 
negligible, we require that $\chi=\gamma_0 E/E_{cr}\ll 1$ \cite{Baier_b_1998}, 
where $\gamma_0$ is the initial Lorentz factor of the electron, 
$E$ is a measure of the amplitude of the electric field $\bm{E}(\bm{\rho})=-\bm{\nabla}\Phi(\rho)=(2\Phi_0/a_s)\bm{\varrho}/[(\eta+\varrho^2+(\eta+\varrho^2)^2]$
in the crystal, and $E_{cr}=m^2/|e|=1.3\times 10^{16}\;\text{V/cm}$ is the critical electric field of QED. 
By employing $E\sim \Phi_M/\rho_c$ as an estimate of $E$, it is $\chi=1.5\times 10^{-5}\varepsilon_0[\text{GeV}]|U_M[\text{eV}]|/\rho_c[\text{\AA}]$.

In the classical regime $\chi\ll 1$ the electron dynamics including RR effects
is described by the LL equation \cite{Landau_b_2_1975}. The LL equation 
for an electron with arbitrary momentum 
$\bm{p}(t)=m\gamma(t)\bm{\beta}(t)$, with $\gamma(t)=\varepsilon(t)/m=1/\sqrt{1-\bm{\beta}^2(t)}$ 
and $\bm{\beta}(t)=\dot{\bm{r}}(t)=d\bm{r}(t)/dt$, reads:
\begin{equation}
\label{LL}
\begin{split}
\frac{d\bm{p}}{dt}=&e\bm{E}+\frac{2}{3}\frac{e^2}{m}\Big\{ e\gamma(\bm{\beta}\cdot\bm{\nabla})\bm{E}+\frac{e^2}{m}(\bm{\beta}\cdot\bm{E})\bm{E}\\
&-\frac{e^2}{m}\gamma^2[\bm{E}^2-(\bm{\beta}\cdot\bm{E})^2]\bm{\beta}\Big\}.
\end{split}
\end{equation}
Here the first two terms of the RR force originate from the Schott
term in the LAD equation whereas the last ``damping'' one corresponds to the Li\'{e}nard
formula. Unlike the first ``derivative'' term, however, the second term of the RR force
is strictly related to the damping one as only their sum ensures that the on-shell 
condition $\varepsilon(t)=\sqrt{m^2+\bm{p}^2(t)}$ is preserved during the electron motion.

Now, we assume that the crystal extends from $z=0$ to $z=L$ and that 
at the initial time $t=0$, the electron's position and velocity are 
$\bm{r}_0=(x_0,0,0)$, with $0<x_0\le\rho_c$, and $\bm{\beta}_0=(0,0,\beta_{z,0})$,
respectively ($\varepsilon_0=m\gamma_0=m/\sqrt{1-\beta^2_{z,0}}$). With these initial 
conditions, due to the symmetry of the potential $\Phi(\rho)$, it is $y(t)=0$ and 
$E_y(\bm{\rho})=0$ along the electron trajectory. Thus, Eq. (\ref{LL}) substantially 
simplifies and only the equation 
\begin{equation}
\label{LL_x}
\frac{d\beta_x}{dt}=-\left(\frac{F_x}{\varepsilon}+\frac{2}{3}\frac{e^2}{m^2}\frac{dF_x}{dx}\beta_x\right)(1-\beta_x^2),
\end{equation}
for $\beta_x(t)$ is needed below, with $F_x(x)=|e|E_x(x,0)$.

If one first neglects RR, the total energy $\varepsilon(t)+U(|x(t)|)$
is a constant of motion. In the ultrarelativistic 
regime $\gamma_0\gg 1$ of interest here and for typical crystal
parameters it results $|\beta_x(t)|\ll 1$, such that $\varepsilon(t)\approx\varepsilon_0[1+\beta_x^2(t)/2]$
(see, e.g., \cite{Artru_1994,Akhiezer_b_1996,Baier_b_1998}). Indeed, energy conservation implies that 
$|\beta_x(t)|\le\sqrt{2|U_M-U(x_0)|/\varepsilon_0}\ll 1$ (recall that 
$|U(\rho)|\sim 100\;\text{eV}$ \cite{Akhiezer_b_1996,Baier_b_1998}). Finally, with the
considered initial conditions, the quantity $\beta_x(t)$ is periodic in time, 
with period $T_0=\sqrt{8\varepsilon_0}\int_0^{x_0}dx/\sqrt{|U(x)-U(x_0)|}$ 
and angular frequency $\omega_0=2\pi/T_0$ \cite{Akhiezer_b_1996}.

\section{Analytical results}
\begin{figure}
\begin{center}
\includegraphics[width=\columnwidth]{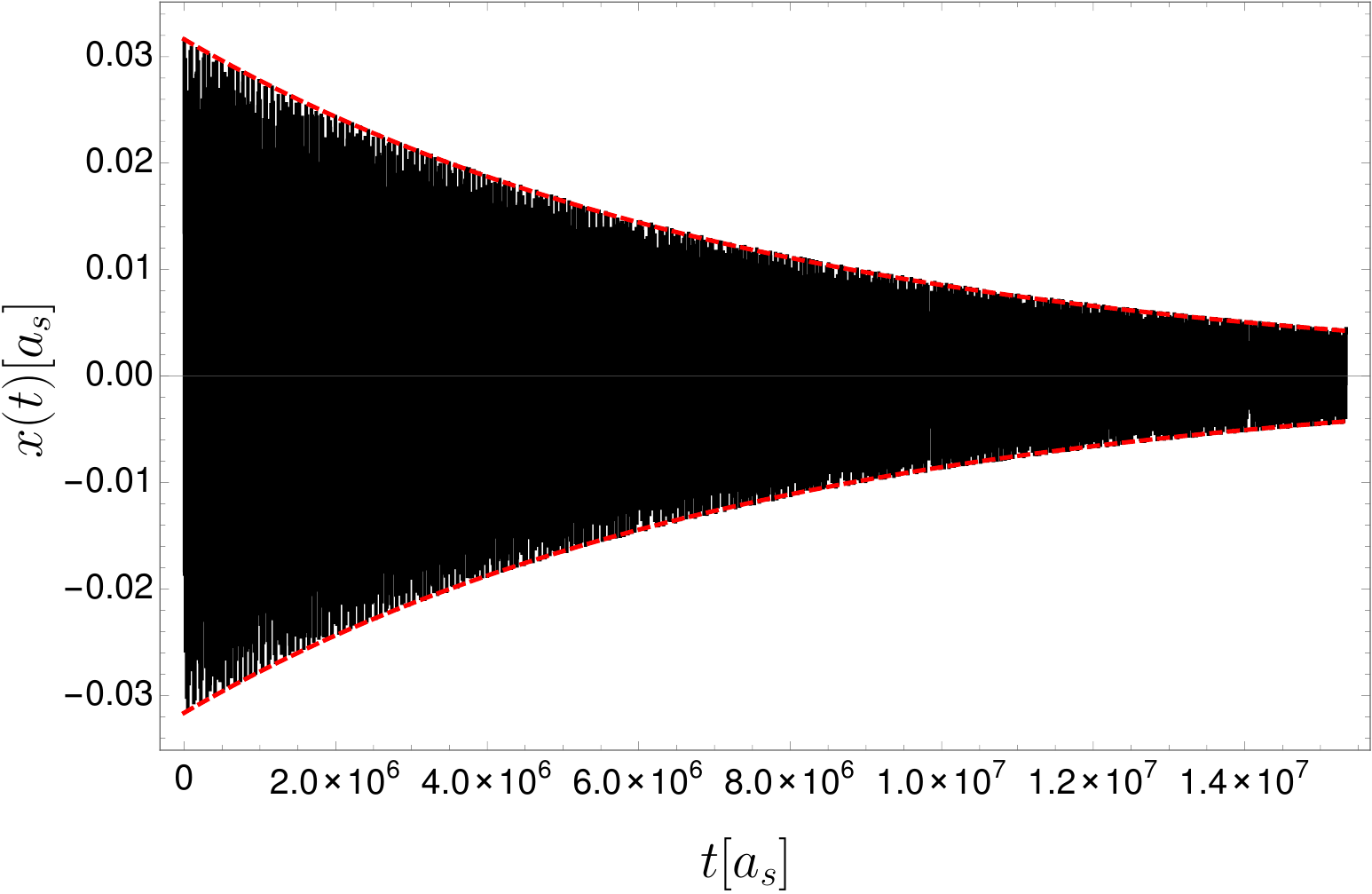}
\caption{(Color online) The rapidly oscillating electron's coordinate $x(t)$ (continuous black curve) 
and the analytical expression $x_0\exp(-t/\tau_d)$ of the envelope (dashed red curve), 
for numerical parameters given in the text.}\label{x_t}
\end{center}
\end{figure}

\begin{figure}
\begin{center}
\includegraphics[width=\columnwidth]{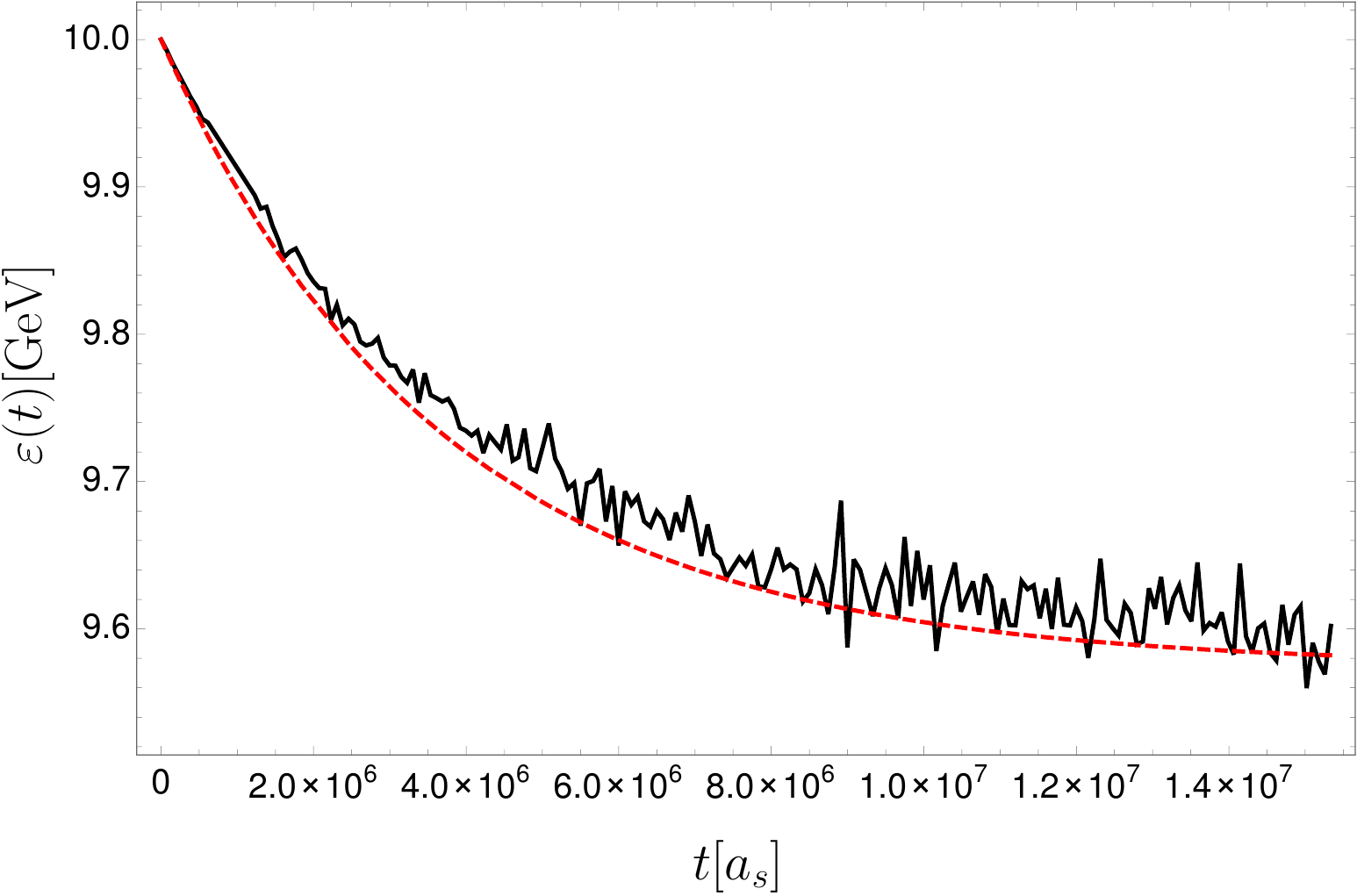}
\caption{(Color online) Time evolution of the electron energy from a numerical 
integration of Eq. (\ref{LL}) (continuous black curve) and according to 
Eq. (\ref{Energy}) (dashed red curve), for the same parameters as in Fig. \ref{x_t}.}\label{gamma_t}
\end{center}
\end{figure}

The considerations above based on the single-string approximation allow us to evaluate 
the effects of RR on the electron dynamics analytically. In fact, as it can be verified 
\textit{a posteriori}, it is safe to assume that $|\beta_x(t)|\ll 1$ and that 
$\beta_z(t)\approx 1$ also including RR. Thus, by multiplying Eq. (\ref{LL}) by $p_x(t)$ and
by neglecting corrections proportional to $\beta_x^2(t)\sim |U_M|/\varepsilon_0$,
it is easy to prove that (see also \cite{Landau_b_2_1975})
\begin{equation}
\label{epsilon}
\varepsilon(t)=\frac{\varepsilon_0}{1+(2/3)
\alpha(\gamma_0/m^3)\int_0^t dt'F^2_x(x(t'))},
\end{equation}
where the integral is performed along the electron trajectory.
In order to get an analytical insight on the motion of the electron, 
we assume here that $|x(t)|\ll a_s\sqrt{\eta}$,
such that $F_x(x)\approx F_0x/a_s\sqrt{\eta}$ and $dF_x(x)/dx\approx F_0/a_s\sqrt{\eta}$, where $F_0=|e|E_0=2|U_0|/a_s\sqrt{\eta}$, with $U_0=e\Phi_0$ ($U_0=-29\;\text{eV}$ for diamond). Equation (\ref{LL_x}) with $1-\beta_x^2(t)\approx 1$ and Eq. (\ref{epsilon})
 show that the electron dynamics along the $x$ direction is characterized by three
time scales: one, $T_0\approx 2\pi /\sqrt{F_0/\sqrt{\eta}\varepsilon_0 a_s}$, proper of the Lorentz dynamics 
and two additional, 
\begin{align}
\tau_s=\frac{6}{\alpha}\frac{\eta}{\gamma_0}\left(\frac{E_{cr}}{E_0}\right)^2\left(\frac{a_s}{x_0}\right)^2\lambda_C, &&\tau_d=\frac{3}{\alpha}\sqrt{\eta}\frac{E_{cr}}{E_0}a_s
\end{align}
introduced by RR and corresponding to the term
containing $F^2_x(x)$ in Eq. (\ref{epsilon}) and 
to the one proportional to $dF_x(x)/dx$  in Eq. (\ref{LL_x}), 
respectively ($\lambda_C=1/m=3.9\times 10^{-3}\;\text{\AA}$ is the Compton wavelength). Now, it is $T_0[\text{\AA}]=1.4\times 10^5a_s[\text{\AA}]\sqrt{\eta\varepsilon_0[\text{GeV}]/|U_0[\text{eV}]|}$, $\tau_s[\text{\AA}]=7.0\times 10^{12}\,\eta^2 a_s[\text{\AA}]^4/(\varepsilon_0[\text{GeV}]U_0[\text{eV}]^2x_0[\text{\AA}]^2)$, and $\tau_d[\text{\AA}]=2.7\times 10^{10}\,\eta a_s[\text{\AA}]^2/U_0[\text{eV}]$, thus for a typical initial energy of $\varepsilon_0=10\;\text{GeV}$
and for $x_0=0.2\,a_s\sqrt{\eta}$ in diamond, it results $\tau_d/\tau_s\approx 0.044$ and $T_0/\tau_d\approx 23\,T_0/\tau_s=1.7\times 10^{-3}$. This suggests to solve Eq. (\ref{LL_x}) by employing the method of
separation of time scales, which provides $x(t)\approx x_0\exp(-t/\tau_d)\cos(\varphi(t))$,
where $\varphi(t)=\int_0^tdt'\omega_0(t')$, with $\omega_0^2(t)=F_0/\sqrt{\eta}\varepsilon(t)a_s$, and
\begin{equation}
\label{Energy}
\varepsilon(t)\approx\frac{\varepsilon_0}{1+(\tau_d/\tau_s)[1-\exp(-2t/\tau_d)]}.
\end{equation}
An alternative derivation of this equation can be obtained
starting from the observation that the momentum $d\bm{P}_r$ 
and the energy $d\mathcal{E}_r$ of the radiation emitted during a time 
$dt$ are related by (see \cite{Artru_2001,Wiedemann_b_2002}) $d\bm{P}_r = \bm{\beta}(t)d\mathcal{E}_r$
and that energy and longitudinal momentum conservation imply
that $d\varepsilon = -d\mathcal{E}_r$ and $dp_z = -dP_{z,r}$ (see in particular 
\cite{Artru_2001,Wiedemann_b_2002} and also \cite{Teitelboim_1969,Teitelboim_1970,Sokolov_2009b} for additional details).
In Fig. \ref{x_t} and Fig. \ref{gamma_t} we show a numerical example 
for diamond indicating the validity of the analytical estimation for 
$x(t)$ and for $\varepsilon(t)$ in Eq. (\ref{Energy}) in comparison with a numerical integration of Eq. (\ref{LL}). 
The initial energy of the electron is $10\;\text{GeV}$, the initial position is $x_0=0.2\,a_s\sqrt{\eta}$, 
and the final time corresponds to a crystal thickness of $0.55\;\text{mm}$ (see also below). 
The above numerical example only aims at showing the validity of our approximated analytical 
treatment and it has to be pointed out that for the used numerical parameters quantum effects 
in the transverse motion of the electron could not be neglected (as it will be clear below, 
the electrons initially so close to an atomic string do not significantly contribute to 
the average emission spectra measured in experiments).
We have ensured in the above numerical example that the trend shown in Fig. \ref{x_t}, 
with RR ``focusing'' the electron's transverse motion to amplitudes much smaller 
than $a_s\sqrt{\eta}$, occurs for all allowed $x_0\le \rho_c$.

\section{Numerical results}
The above considerations provide an analytical insight on the effects of 
RR but hold under the assumption that the crystal potential 
can be approximated by the expression in Eq. (\ref{eq:potential}). 
Below, we will investigate numerically the effects of RR on the emission spectra 
of electrons crossing a diamond crystal along the axis $\langle 100\rangle$, 
with the crystal field being represented more realistically than above 
by a periodic replica of the Doyle-Turner potential \cite{Doyle_1968}. 
Considering the distribution of the strings along the $x\text{-}y$ plane perpendicular
to the axis $\langle 100\rangle$ of diamond, 
at each instant we have included the effects of the 16+25=41 strings within
a square centered on the string closest to the electron. In order to obtain results more easily comparable 
with experimental results, the reported single-particle spectra result 
from the average over 200 electrons all with the same incoming momentum 
(along the $z$-direction) and energy $\varepsilon_0=30\;\text{GeV}$, 
and uniformly distributed over the cell $-a/4\le x,y\le a/4$, with 
$a=3.57\;\text{\AA}$ being the diamond lattice constant.
Now, RR effects are clearly larger for thicker crystals. However, 
an upper limit to ``meaningful'' values of the crystal thickness is 
set by the dechanneling, i.e., by the fact that, due to multiple Coulomb 
scattering with the atoms in the crystal, the transverse amplitude of 
the electron motion increases and, after a certain distance $l_d$ 
(dechanneling length), the electron leaves the ``channel'' generated 
by a single atomic string \cite{Akhiezer_b_1996,Baier_b_1998}. The term 
``meaningful'' above thus refers to the fact that for a crystal thickness 
much larger than $l_d$, the electron will not anyway emit channeling radiation 
after a distance of the order of $l_d$. An order-of-magnitude estimate of the dechanneling 
length $l_d$ for an electron initially propagating along the atomic 
string is given by $l_{d}=(\alpha/4\pi)(|U_M|\gamma_0/m)X_{0}$, where 
$X_{0}=[4Z^2\alpha^3 n\lambda_C^2\log(183 Z^{-1/3})]^{-1}$, with $Z$ being 
the crystal atomic number and $n$ its atomic density, is the radiation 
length in the amorphous case \cite{Baier_b_1998}. 
In order to implement the effects of multiple scattering and of 
dechanneling on the electron motion, we started from the kinetic equation 
describing the evolution of the transverse velocity with respect to time, 
which can be approximated as a Fokker-Planck equation with diffusion 
coefficient $D/4=\overline{\beta_{\perp}^2}/4L$, where 
$\overline{\beta_{\perp}^2}=(4\pi/\alpha)L/\gamma_0^2 X_0$ and $L$ is the 
thickness of the crystal \cite{Akhiezer_b_1996,Baier_b_1998} (note that
the dechanneling length corresponds to the thickness obtained by equating the quantity
$\sqrt{2\overline{\beta_{\perp}^2}}$ with the Lindhard critical angle 
$\theta_c=\sqrt{2|U_{M}|/\varepsilon_0}$). Based on 
the equivalence between the Fokker-Planck kinetic equation and a single-particle 
stochastic equation \cite{Gardiner_b_2009}, we have added the stochastic 
term $d\bm{\beta}_{\perp,s}=\sqrt{D/2}\,d\bm{S}_t$ to the equation of motion 
for the transverse velocity $\bm{\beta}_{\perp}$ of the electron, where 
$\bm{S}_t$ is the vector stochastic variable corresponding to the Wiener 
process and having the dimension of the square root of time \cite{Gardiner_b_2009}. 
Each spectrum has then been obtained by averaging over five spectra, every 
one being obtained with an independent sequence of random numbers corresponding 
to the stochastic variable $\bm{S}_t$. We point out that the diffusion 
coefficient $D$ corresponds to the amorphous case, whereas the electrons 
within a disk of radius $u_{\perp}\approx 0.04\;\text{\AA}$ and centered on a string, 
with $u_{\perp}$ being the average thermal vibration amplitude on the plane
perpendicular to the string, would see a relatively high 
nuclear density and would dechannel at distances significantly smaller than 
$l_d$. In order to include the effect of the higher nuclear density in the vicinity
of the strings, we have followed Ref. \cite{Beloshitskii_1982} and we have multiplied
the diffusion coefficient $D$ by the enhancing factor $P(\bm{r})=(s/\pi u_{\perp}^2)\exp(-\rho^2/u_{\perp}^2)$, where $s=1/nd=1.6\;\text{\AA}^2$ is the area for each string, with $n=1.77\times 10^{23}\;\text{cm$^{-3}$}$ for diamond and $d=a=3.57\;\text{\AA}$ being the distance between two atoms in a string, and where $\rho$ is the distance of the electron from the closest string.

\begin{figure}[t]
\includegraphics[width=1\columnwidth]{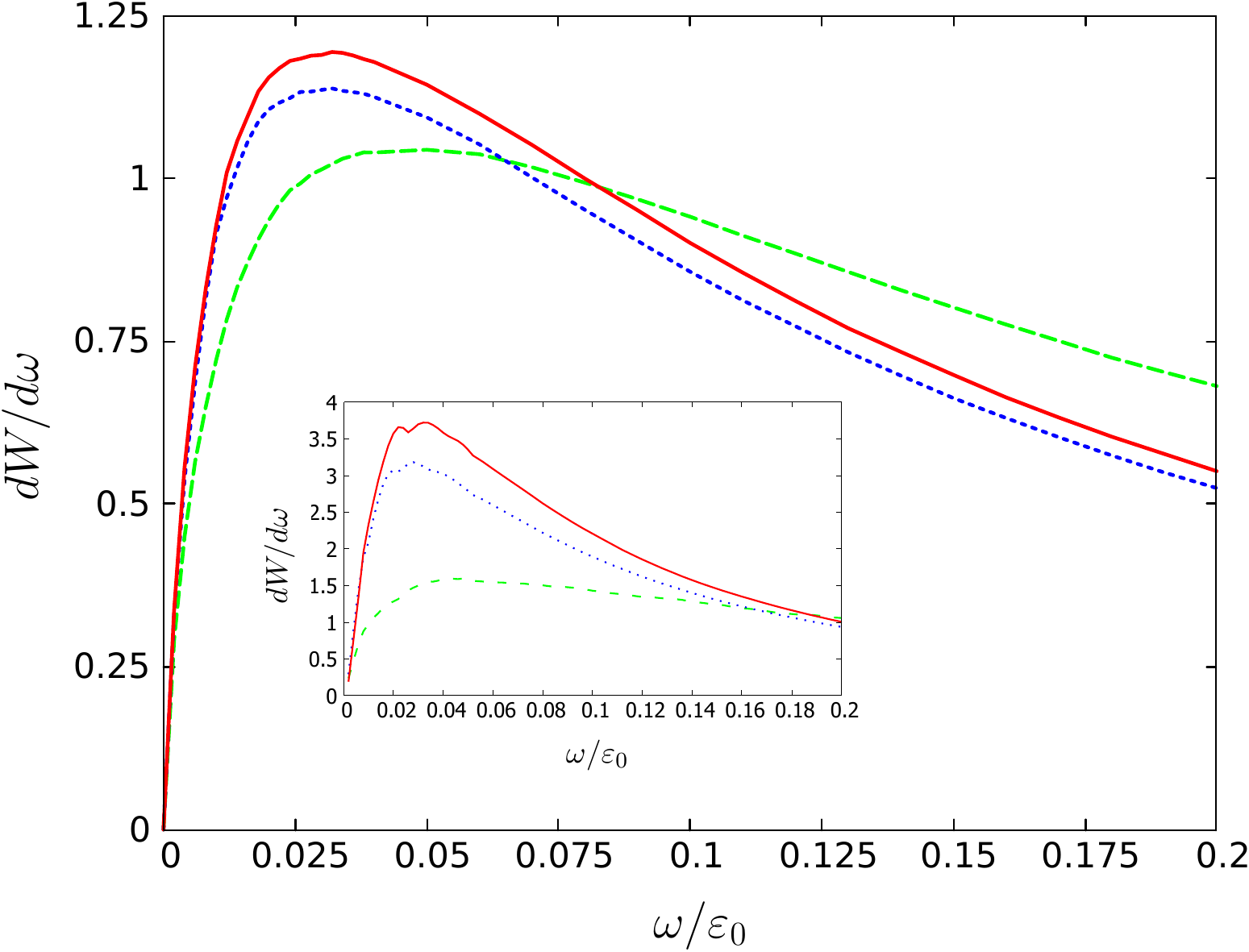}
\caption{(Color online) Radiation energy spectra for numerical parameters given in the 
text without RR (dashed green curve), with RR excluding the derivative term 
in the LL equation (dotted blue curve), and with RR including the derivative 
term in the LL equation (continuous red curve). The inset shows the corresponding 
spectra without multiple scattering.} \label{fig:Spectra}
\end{figure}

In Fig. \ref{fig:Spectra} three single-electron energy spectra $dW/d\omega$ 
are shown as a function of $\omega/\varepsilon_0$, with $\omega$ being the 
emitted radiation angular frequency, and for a crystal thickness of 
$L\approx l_d/3=0.55\;\text{mm}$. In order to test specifically the 
importance of the derivative term in the LL equation (\ref{LL}), we show 
the spectrum without RR terms (dashed green curve), with RR terms except 
the derivative one (dotted blue curve), and with all RR terms (continuous 
red curve). The inset shows the corresponding spectra without the inclusion of multiple scattering.
The spectra are calculated by integrating the differential spectrum \cite{Jackson_b_1975}
\begin{equation}
\label{Spectrum}
\frac{dW}{d\omega d\Omega}=\frac{e^2}{4\pi^2}\left\vert\int_{-\infty}^{\infty} dt\frac{\bm{n}\times[(\bm{n}-\bm{\beta})\times \dot{\bm{\beta}}]}{(1-\bm{n}\cdot\bm{\beta})^2}e^{i\omega(t-\bm{n}\cdot\bm{r})}\right\vert^2
\end{equation}
with respect to the solid angle $\Omega$ along the observation direction $\bm{n}$ (see also \cite{Wistisen_2014} for details) and by integrating numerically either the Lorentz equation or the LL equation along the whole electron trajectory. The Lorentz equation corresponds to the dashed green curve and the LL equation to the dotted blue curve (without the derivative term) and to the continuous red curve (with the derivative term). For the considered numerical parameters, the local constant crossed field approximation \cite{Baier_b_1998}, which requires $\langle K(t)\rangle\gg 1$, where $\langle K(t)\rangle=\sqrt{2\langle\gamma^2(t)\beta_{\perp}^2(t)\rangle}$ is the average Free-Electron Laser (FEL) parameter, with $\langle f(t)\rangle=L^{-1}\int_0^Ldtf(t)$, cannot be applied here. In fact, as it can be seen in Fig. 4a, where for the sake of simplicity multiple scattering is ignored, it turns out that for the above numerical parameters it is $\langle K(t)\rangle\lesssim 1$.
\begin{figure}[t]
\includegraphics[width=1\columnwidth]{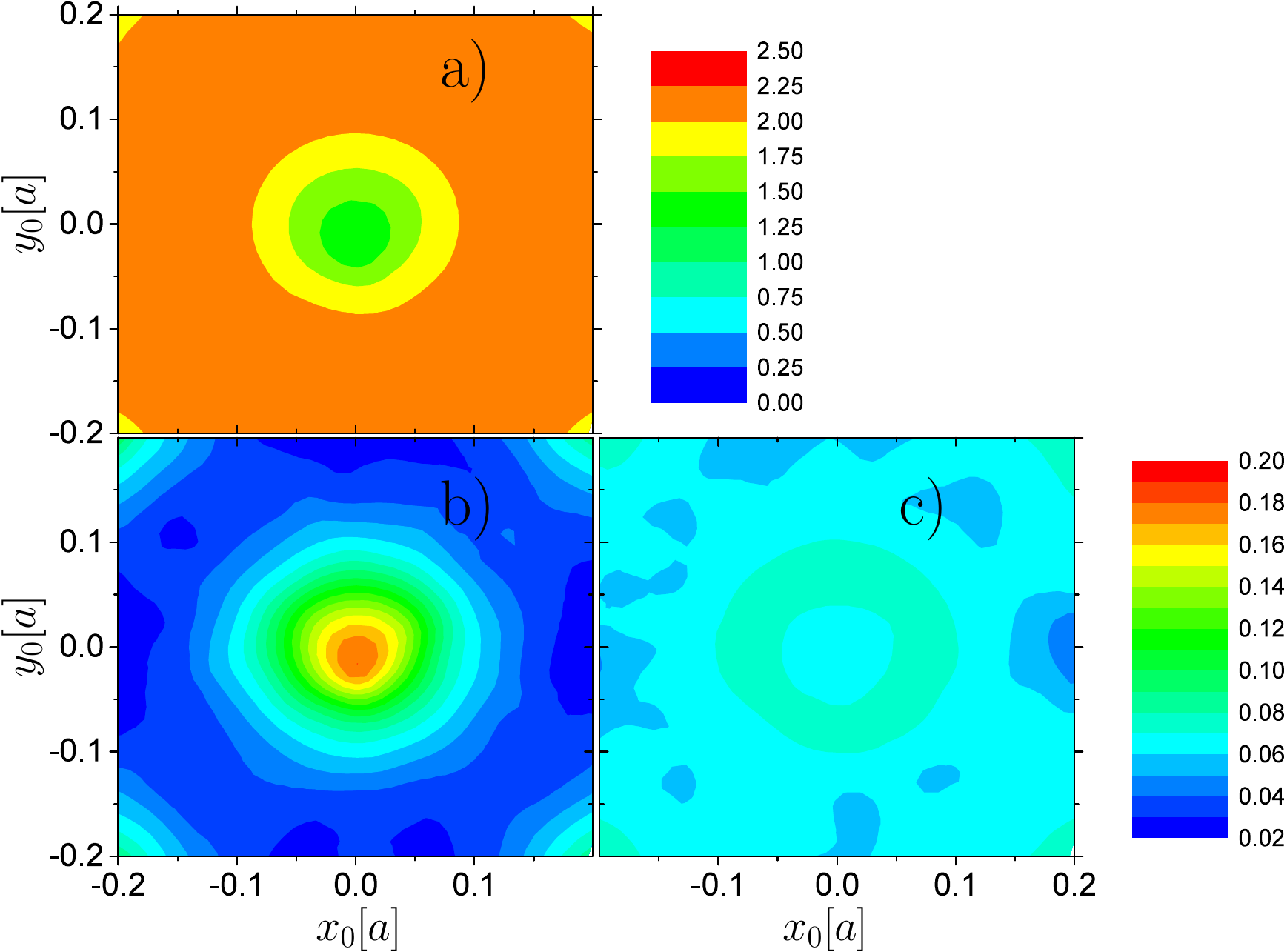}
\caption{(Color online) The average FEL parameter $\langle K(t)\rangle$ with RR (part a)) and the average electric field $\langle\chi(t)\rangle$ experienced by the electron in units of $E_{cr}$ without (part b)) and with (part c)) RR as functions of the initial transverse coordinates of the electron (a crystal atomic string is at the origin of the coordinates). The numerical parameters are the same as in Fig. \ref{fig:Spectra}.}
\label{K_av_chi_av}
\end{figure}
Going back to Fig. \ref{fig:Spectra}, the main effect of RR is to increase the radiation yield at low frequencies and the derivative term in the LL equation enhances this effect. The resulting lowering of the average emitted radiation frequency can be understood qualitatively as RR effects tend to reduce the electron energy and the average electric field experienced by the electron. The reduction of the average electric field experienced by the electron, corresponding to the parameter $\langle\chi(t)\rangle$ is not obvious because RR also induces a cooling effect in the transverse motion (see, e.g. Fig. 1) which, for some values of the initial electron's position, might let the electron spend more time in regions where the electric field is large. In Figs. 4b and 4c we show the quantity $\langle\chi(t)\rangle$ as a function of $x_0$ and $y_0$ without RR (Fig. 4b) and with RR (Fig. 4c)) and, again, ignoring multiple scattering for simplicity (as it can be seen from Fig. \ref{fig:Spectra} multiple scattering does not alter qualitatively the effects of RR). Although, for some values of $x_0$ and $y_0$ RR indeed induces an increase of $\langle\chi(t)\rangle$, for those initial conditions closer to the atomic string and corresponding to the largest values of $\langle\chi(t)\rangle$, RR induces a reduction. It is worth mentioning here that RR effects are most important at the peak of the emission spectrum corresponding to photon energies of the order of $0.025\,\varepsilon_0$, where quantum effects are safely negligible. The enhancement of RR effects due to the derivative term in the LL equation can be qualitatively understood going back to the simplified model in Section 2 and by noticing that the derivative $dF_x/dx$ is largest at small $x$, where $dF_x/dx>0$, such that the corresponding term in Eq. (\ref{LL_x}) acts as an additional ``cooling'' term.

It is also worth observing that multiple scattering with the nuclei 
tends to increase the transverse electron energy. As expected, this 
``heating'' effect, on the one hand decreases the overall emission yield
and, on the other hand, also suppresses the cooling effect due
to RR (see Fig. \ref{fig:Spectra}). However, Fig. \ref{fig:Spectra} 
shows that for the chosen numerical parameters, the effects of RR 
and in particular of the derivative term in the LL equation are still sizable,
although detecting the latter experimentally may prove to be challenging.

Finally, on the one hand, our numerical model including the effect of multiple
atomic strings on the electron motion takes into account automatically
the radiation by dechanneled electrons in the corresponding potential.
On the other hand, it can be checked that the contribution $dW_{IB}/d\omega$
of incoherent bremsstrahlung to the emission spectrum for the numerical
example in Fig. \ref{fig:Spectra}
is negligible. In fact, starting from the Bethe-Heitler cross section
(see e.g. Eq. (27) in \cite{Uggerhoj_2005}), it can be seen that in the
region $\omega\ll\varepsilon_0$, the function $dW_{IB}/d\omega$ is approximately
constant and
\begin{equation}
\frac{dW_{IB}}{d\omega}\approx\frac{16}{3}Z^2\alpha^3 n\lambda_C^2L\log(183 Z^{-1/3})=\frac{4}{3}\frac{L}{X_0}.
\end{equation}
By plugging the numerical parameters corresponding to the plots in Fig. \ref{fig:Spectra}, one
obtains that $dW_{IB}/d\omega\approx 5\times 10^{-3}$. In addition, we have ensured that by also accounting for the higher nuclear density experienced by the electrons close to the atomic strings at channeling than in an amorphous medium, the effect of incoherent bremsstrahlung is still negligible. In fact, following \cite{Beloshitskii_1982}, this amounts in multiplying the quantity $dW_{IB}/d\omega$ by the enhancing factor $\langle P(\bm{r}(t))\rangle$. We have ensured that, by including the effects of RR and of multiple scattering, the quantity $\langle P(\bm{r}(t))\rangle$ is typically smaller than 10 for almost all initial conditions as in Fig. 4 and that its average value with respect to the initial conditions is typically less than 3.

\section{Experimental considerations}
\begin{figure}
\begin{center}
\includegraphics[width=\columnwidth]{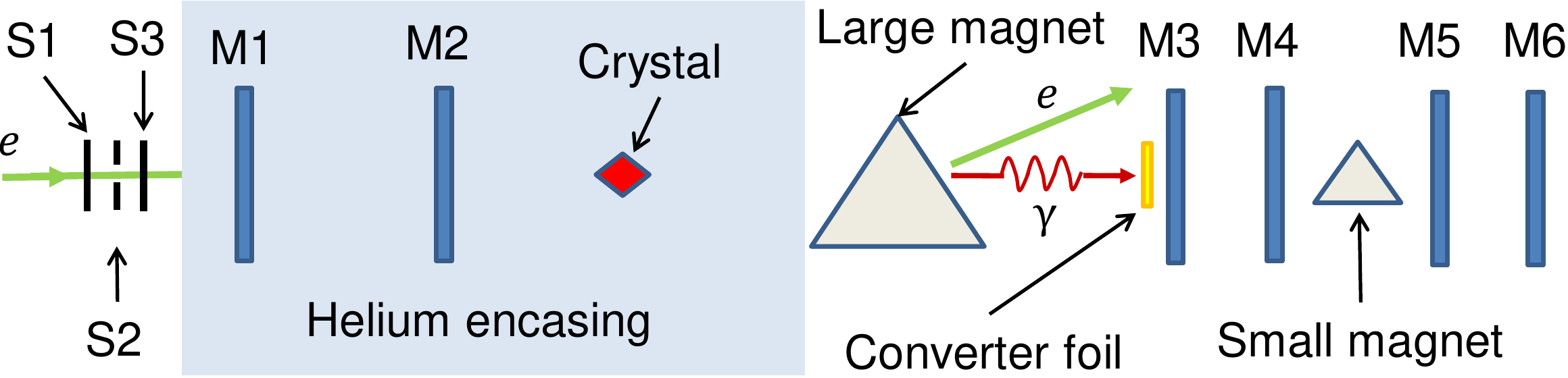}
\caption{Sketch of a possible experimental setup (top view). S1-S3 denote scintillators and M1-M6 denote position-sensitive MIMOSA detectors \cite{Winter2010192}.}\label{fig:expfig}
\end{center}
\end{figure}

Measurement of the spectra in Fig. \ref{fig:Spectra} is possible using a setup as shown in Fig. \ref{fig:expfig}.
After passing the scintillators S1-S3, the electrons go through two position-sensitive MIMOSA detectors M1 and M2 \cite{Winter2010192} encased in Helium to reduce multiple scattering, in order to determine their incoming angle \cite{Kirs01}. By deflecting the charged particles outgoing from the crystal via the large magnet, only the emitted photons hit a converter foil to produce electron-positron pairs. By measuring the energy of the pairs employing the small magnet, the energy of the photons can be determined. The case considered here of electrons initially moving along the atomic string is a reasonable approximation as long as the electrons impinge with angles to the atomic string on a scale of order of or smaller than the Lindhard critical angle $\theta_{c}$. Electrons with an angular divergence comparable to $\theta_c$ can indeed be achieved at the CERN SBA \cite{SBA}. The spectrum including RR in the inset in Fig. \ref{fig:Spectra} corresponds to each electron emitting approximately 4.4 photons capable of pair production in the converter foil. In order to avoid pileup and obtain single-photon spectra, the converter foil should have correspondingly a thickness smaller than about one fifth of the radiation length. In the region around the peak of the red curve in the inset in Fig. \ref{fig:Spectra} where $dW/d\omega>1$ about 3.5 photons are emitted. In order to resolve the peak in 200 bins with $10^{4}$ counts in each bin corresponding to an uncertainty of $1\%$, which would allow to discriminate among the three higher peaks of the curves in the inset in Fig. \ref{fig:Spectra}, would thus require about $2.9\times 10^6$ electrons. At the CERN SBA a rate of 2000 electrons per minute can be achieved implying a measurement time of about 24 hours.

\section{Conclusions}
In conclusion, we have demonstrated that the predictions of the
LL equation can be feasibly tested experimentally by measuring the 
channeling radiation emitted by ultra-relativistic electrons
impinging onto a diamond crystal slab. The required experimental 
conditions are available at the CERN SBA beamlines. Most importantly, the effects of the derivative term in the LL equation are
shown to affect the emission spectra much more than
in previous proposals based on intense lasers fields although the measurability of such effects
may be challenging. In this respect, we point out that the present
one represents the first investigation on testing the LL equation
in aligned crystals and a more complete and quantitatively precise
study would include other effects than those already considered
here as the incidence angle of the electrons or the efficiency 
and the resolution of the detectors.

\section*{References}

\bibliographystyle{elsarticle-num}

\end{document}